\documentclass[12pt]{article}
\usepackage{fullpage}		%Give a standard 1 inch margins on all sides
\usepackage{amsmath,amssymb,float}
\usepackage{graphics}
\usepackage{subfigure}
\usepackage{epsfig}
\usepackage{lineno}
\begin{document}
\title{Weinberg Angle and Integer Electric Charges of Quarks}
\author{Martin A. Faessler \footnote{Supported by the DFG Cluster of excellence 
Origin and Structure of the Universe}\\
 Ludwig-Maximilians-University, Physics Department, 80799 Munich, Germany
 }
\date{\today}
\maketitle

\abstract{\noindent 
%Most measurements cannot distinguish whether electric charges of quarks are fractional, 
%or appear to be fractional as a result of averaging over  integer electric charges. 
%Yet, processes where the coupling of the gauge boson(s) involve the square of the electric charge 
%of quarks  can distinguish between the two alternatives. 
%Such is the case for the amplitude describing the orthogonality between photon and Z boson. 
Orthogonality between  $\gamma$ and $Z$ boson involves the Weinberg angle and a scheme for assignments 
of electric charge and weak isospin to leptons and quarks coupling to  $\gamma$ and $Z$.
The Han-Nambu scheme,  with integer electric charges for quarks, satisfies $\gamma Z$  orthogonality 
with $sin^2(\theta _W)= 0.25 $ in leading order. 
%Assuming quarks have integer electric charge allows to derive the Weinberg angle in agreement
%with its experimental value. 
Experimental results for  photon-photon fusion into $c \bar{c}$ and  $b \bar{b}$
pairs provide further support for assigning integer electric charges to quarks.  
%$x_{Bj}$, $Q^2$, $W^2$  
%$\langle k_{\perp}^2 \rangle$ with $Q^2$ is observe
} 
%
%\maketitle
% useful symbols ^  |  
%
%\section
\\
\\

\section{Introduction}
Whether quarks have fractional or integer electric charges is not an issue of present particle physics 
but has been discussed extensively in the  past, e.g. 
\cite{HanNambu1965} - 
%\cite{PatiSalam1973},
%\cite{ChengWilczek1974},
%\cite{LipkinRubinstein1978}, 
\cite{IijimaJaffe1981}.

The present Standard Model \cite{GSW} of particle physics 
(see  \cite{Textbooks} %Perkins, Griffith, Okun,Martin and Shaw, Close, 
and \cite{Pdg2012}) assumes that quarks have fractional charges.  
In units of e  they are  
+2/3 for  up, charm and top quarks and -1/3  for down, 
strange and bottom quarks independent of their color, as proposed by
Gell-Mann  \cite{Gellmann1964} and Zweig \cite{Zweig1964} in the framework of flavor-SU(3) 
and later adopted by Fritzsch, Gell-Mann and Leutwyler in the framework of 
color-SU(3) and QuantumChromoDynamics \cite{FritzschGellmann1971}.

As an alternative scheme, Han and Nambu \cite{HanNambu1965} had proposed that quarks have integer charges, 
but different for the three colors.
Here, the fractional charge is the result of the superposition of the
three quarks to a color singlet. 
Table 1 shows the electric charges $Q_{fc}$ according to this scheme for one generation of
quarks, consisting of three up- and three down-like quarks with different strong color charges.

\begin{table}[htdp]
\centering
\begin{tabular}{   |p{3.0cm}|p{2.0cm}p{2.0cm}p{2.0cm}| } \hline

  Color charge $\rightarrow$    & red   &    blue    & green \\
 Flavor $\downarrow$            &       &            &       \\
\hline
     up                         &   +1  &   +1       &    0  \\ 
    down                        &    0  &    0       &   -1  \\
\hline

\end{tabular}
\caption{Electric charge  $Q_{fc}$  in units of e, the absolute value of the electron charge, for the
two quark flavors $f$ = up and down, and three color charges $c$ =  red, blue, and green, according to the Han-Nambu scheme.
}
\label{Table1}
\end{table}

Obviously, for an up quark in a color singlet state 
$| up \rangle$ with the short notation:

\hspace{1cm} $| up \rangle = 
(| up(Q=+1)_{red}  \rangle + |up(Q=+1)_{blue}  \rangle +| up(Q=0)_{green} \rangle ) /\sqrt{3} $\\
the average charge $\langle Q \rangle$  is +2/3 and, correspondingly, for the down quark, -1/3.

It has been realized, also decades ago \cite{LipkinRubinstein1978}, 
that as long as quarks are confined in color singlet
states the direct coupling of photons to quarks measures the average charge and cannot distinguish
between the two alternatives, fractional charges according to the Standard Model or a superposition of
integer charged quarks according to Han-Nambu.

\begin{figure}[t]
      \begin{center}
            \resizebox{0.35\columnwidth}{!}{\includegraphics{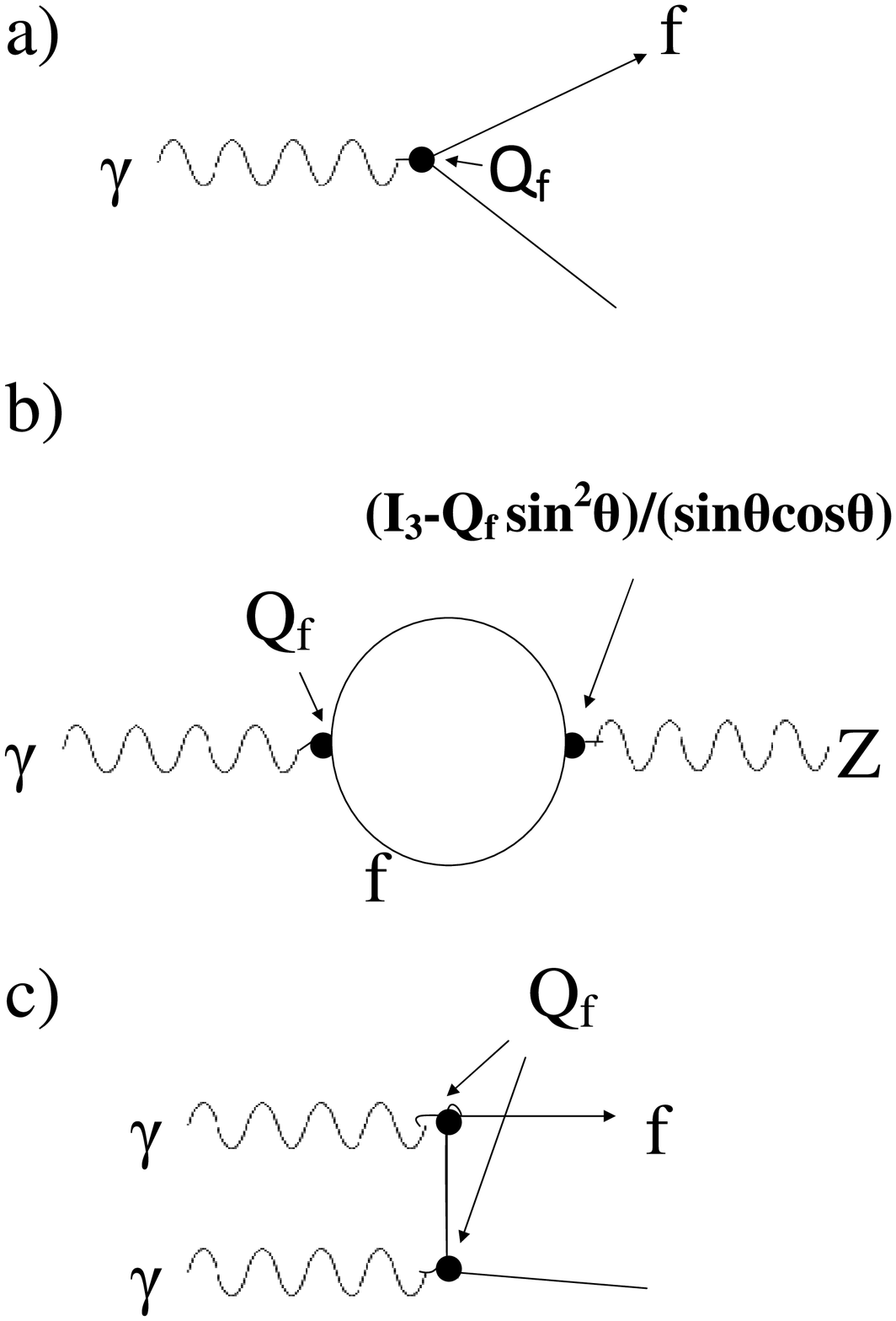} }  %0.5 instead of 0.85
      \end{center}
\caption{Leading order Feynman diagrams for
a) decay of $\gamma$  into  charged fermion-antifermion ($f \bar{f} $);  
b) coupling  of  $\gamma$ and Z boson via internal charged fermion loop; 
c) direct  $\gamma \gamma$ fusion to charged $f \bar{f} $ 
}
\label{Fig1}    
\end{figure}

The strongest experimental evidence for fractional (average) electric quark charges 
comes from $ e^+ e^-$ annihilations into hadrons. 
The standard description is based on the Feynman diagram, see Fig.1a,
where the virtual photon results from the  $ e^+ e^-$ annihilation. 
Firstly, the virtual photon $\gamma$ 
decays into a quark and antiquark  
and subsequently this pair fragments into hadrons. 
The ratio of the cross sections $ \sigma ( e^+ e^- \rightarrow hadrons)$ and 
$  \sigma ( e^+ e^-  \rightarrow \mu^+ \mu^-)$: 
%\hspace{5cm }  
$R = \sigma ( e^+ e^- \rightarrow hadrons) / \sigma ( e^+ e^-  \rightarrow \mu^+ \mu^-)$ %\\ 
displays the emergence of new quarks as a function of the virtual photon invariant mass,
starting from  thresholds dictated  by the effective mass 
of the quark-antiquark pair, and provides a direct
measure of the electric charges of the quarks and the number ($N_c =3$) of different colors. 
The ratio is described by the formula 
$ R =  \sum _f   (\sum _c    Q_{fc} /\sqrt{3} ) ^2 $, 
where the indices $f$ and $c$  indicate the quark flavor  and color charge  
and the sum extends over all quark flavors   
which can be produced at the given virtual photon invariant mass. 
Consider the production of a quark-antiquark pair $f \bar{f} $ of flavor $f$ 
with electric  charge(s) $Q_{fc}$   in the Standard Model  or the Han-Nambu scheme. 
The transition amplitude $T_f$ is proportional to the coupling 
of the photon to the effective electric quark charge, 
the quark being in the color singlet  state, i.e.

 \hspace{3cm } $T_f \propto  (Q_{f red} + Q_{f green} + Q_{f blue} ) /\sqrt{3} = 3 \langle Q_f \rangle /\sqrt{3}$\\
This (partial) amplitude contributes to the cross section ratio $R$ with 
$ (3 \langle Q_f  \rangle / \sqrt{3})^2 = N_c  \langle Q_f \rangle ^2$ . 
The experimental data are in perfect agreement with the assumption of the Standard Model 
up to the highest measured virtual photon masses.
%above the Z boson, below the top quark threshold. 
But they are also in agreement with the Han-Nambu scheme as long as quarks are % supposed to be 
confined to color singlet states.

Another (lost) testing ground for the distinction between fractional and integer quark charges 
is Deep Inelastic Scattering (DIS) of leptons. 
The exchanged virtual photon, now spacelike with imaginary invariant mass, couples in first order
to quarks in color singlet states. 
(This part of DIS is as well described by diagram Fig.1a). 
Thus, DIS as well as $ e^+ e^-  $ annihilation cannot distinguish between the Standard Model 
and the Han-Nambu scheme. 
The proof that the (average) electric charge is fractional for DIS 
can be derived from the ratio of $F_2$ structure functions 
measured in charged electron and muon DIS (photon exchange) on average nucleons 
over that measured in neutrino DIS (W boson exchange). 
This ratio amounts to about $ 5/18 = 1/2 (\langle Q_{up} \rangle^2 + \langle Q_{down} \rangle^2)$.

\section{Estimating  the Weinberg angle}
In order to distinguish between integer and fractional charges of quarks, 
transition amplitudes are required where the
square of the electric charges of quarks enter. 
This has been emphasized  by Witten already in 1977 \cite{Witten1977}.

First consider the diagram shown in Fig.1b. 
It displays the lowest order coupling between $Z$ boson  and photon $\gamma$ through the
internal loop of a fermion $f$. 
The Standard model postulates orthogonality between  $\gamma$  and  $Z$. 

(1) \hspace{4cm }   $  T_{\gamma Z} = \langle Z| \gamma \rangle = 0 $  \\
The amplitude  $  T_{\gamma Z} $ implies the summation over all fermions, quarks and leptons. 
The product of coupling constants of each fermion to   $\gamma$  and  $Z$  enter. 
The coupling  to $Z$ depends on % is mediated  by its 
electric charge and  third component  $I_3$ of the weak isospin of the fermion. 
The left-handed fermions have, according to the Standard Model,
$I_3 = +1/2$ (neutrinos  and up quarks), $I_3 = -1/2$ (charged leptons and down quarks). 
Right-handed charged leptons, up and down quarks have  $I_3 = 0$. 
Thus, for a given internal fermion $f$ (with color charge $c$, in case of quarks), 
the coupling  $C_{\gamma  Zfc}$ between photon and Z boson 
is, according to  the Standard Model, omitting a common factor $1/ (sin(\theta _W )cos(\theta _W))$ :

(2)   \hspace{3cm }   $C_{\gamma Z fc} =  Q_{fc} \cdot [I_{3f} -Q_{fc} \cdot sin^2(\theta _W)]$ \\ 
%\noindent  
with the Weinberg angle $\theta_W$. 

Let us assume, all fermions are massless. 
And restrict the consideration to the first family of known fermions,
since the additional families are repetitions of the first with respect to the quantum numbers relevant
for the coupling to  $\gamma $ and $Z$.  
Summing over all fermions, leptons and quarks, in order to have orthogonality (1)
simplifies to requiring that the sum of the couplings  $C_{\gamma  Zfc}$  vanishes.
There are 14 family members  which contribute: the left and right handed electron, 
$e⁻_R, e⁻_L $ and  quarks 
$up_{Rc}, up_{Lc}, down_{Rc}, down_{Lc}$,  with the three colors $c = red, blue, green$.  
Summing over color and handiness, the partial contributions  $C_{\gamma Z f} $ of electron, up and down quarks to 
the orthogonality condition are obtained. 
They are of course identical for the electron, in both schemes, Standard Model (SM) and Han-Nambu (HN):

\hspace{3cm }  $   C_{\gamma Z e}  = [1 -4 sin^2(\theta _W)]/ 2 $. \\
For the quarks, the results differ in both schemes. The SM  obtains: 

\hspace{1cm }   $ C_{\gamma Z up}   = [1 - (8/3) sin^2(\theta _W)]    $   \hspace{1.5cm } 
    $ C_{\gamma Z down} = [1 - (4/3) sin^2(\theta _W)]/2  $\\
For the HN scheme one obtains: 

\hspace{1cm }   $  C_{\gamma Z up}  = 2[1 - 4 sin^2(\theta _W)]  $   \hspace{2cm } 
    $  C_{\gamma Z down} = [1 - 4 sin^2(\theta _W)]  $\\
Requiring the internal loops of quarks to be color singlets, all the quark contributions 
have to be multiplied  with a factor $1/ \sqrt{N_c} = 1/  \sqrt{3}  $ for both schemes.   

As a final step, the contributions  of electron and the two quark flavours are added up to extract the 
Weinberg angle from the orthogonality relation (1)  (under the assumption of massless fermions): 
% and restriction to one generation):     

(3) \hspace{3cm } $T_{\gamma Z} \propto \sum _f  C_{\gamma Z f}  = 0$ \\ 
%Therefore, (1) and (2)  allow to determine the Weinberg angle with the following relation
%
%(4)   \hspace{3cm } $sin^2(\theta_W) = \sum _{fc} (Q _{fc} \cdot  I_{3f }) / \sum _{fc} Q^2_{fc} $ \\
%Assuming for the three known  generations of fermions a repetition of all quantum numbers 
%apart from flavor, and furthermore assuming that all members of the
%three generations are known at present, it is sufficient to consider one generation of fermions.
%The multiplet of fermions of one generation incorporating quarks according to the Han-Nambu scheme 
%has the following properties as compared with the Standard Model multiplet:
%
%The total sum of electric charges is 0 for both alternatives, 
%integer and fractional electric charges of individual quarks. 
%Hence   $\sum Q_f \cdot I_{3f} = 2$, in both cases. 
%According to the Standard Model $\sum Q^2_f = 16/3$, 
%hence, for the Weinberg angle one obtains
The SM scheme leads to $ sin^2(\theta _W) = 3/8 = 0.37$   or  $ sin^2(\theta _W) = 0.35$,
where the latter value is obtained applying the factor  $1/ \sqrt{N_c}  $  to the quark amplitudes.
 
According to the HN scheme, the multiplet of one family with integer charges yields 

\hspace{4cm } $ sin^2(\theta _W) = 1/4 = 0.25 $, \\
independent of the color singlet assumption. 
The latter value is  much closer to  the best present value of the Weinberg angle \cite{Pdg2012} 
determined from many experimental results and extrapolated to the  $Z$ mass $M_Z$ 
in the modified minimal subtraction scheme:  %, just two standard deviations away :

\hspace{4cm } $sin^2(\theta _W (M_Z)) = 0.23116  \pm 0.00012$ 

In the above derivations the mass of all fermions has been neglected. 
To calculate  $  T_{\gamma Z} $,  each  coupling factor   $C_{\gamma f  Z} $ has to be multiplied  
with a weight $( \propto 1/ m^2_f ) $, which depends on the 
mass   $ m_f $  of  the fermion in the internal loop of the diagram Fig. 1b.   
%and the invariant mass of the external $\gamma$ and $Z$ line in diagram 

However,  under the assumptions of the HN scheme, 
all three charged fermions of one generation contribute to the sum with a product 
containing the same  factor $[1 - 4 sin^2(\theta _W )]$.
Hence, lepton, up and down quark,  separately satisfy 
the orthogonality condition, independent of their mass, for a Weinberg angle $sin^2(\theta _W) = 1/4 $. 
Even more generally: Every single colored  fermion with non-zero integer electric charge and  
with the same assignment rules  for the signs of $ I_3 $ and $Q$ 
%electric and weak charge quantum number 
as those valid for  charged  fermions in the Standard Model: 
equal signs of   $ I_3 $ and $Q$ for the left handed, 
%and $I_3 = 0$ and $Q= -1$ for the right handed leptons, 
separately satisfies the orthogonality (1). 
%It appears that a theory following these rules has to postulate   $sin^2(\theta _W) = 1/4 $ 
%in order to satisfy $ \gamma  Z $ orthogonality.

Adopting the HN scheme,  orthogonality is satisfied for a Weinberg angle with $sin^2(\theta _W) = 1/4 $, 
since the $Z$ couples to a purely  axial and  $  \gamma  $ to a purely vector 
fermion current.
In the above derivation the internal loop with a $W$ gauge boson instead of fermion 
has been neglected. 
This is only justified by the empirical   fact that the mass of $W$ is large compared to 
the masses of fermions of the first generation. 

%This appears to be a strong argument that nature should prefer integer electric quark charges. 
In the absence of a better prediction for the Weinberg angle based on the SM scheme this may be considered 
as support for the HN scheme, i.e. integer electric charges of quarks.

%Case 2: Two-photon coupling to  fermions
\section{Two-photon physics}
Two photon interactions have been extensively studied at the Large Electron Positron Collider LEP. 
Of particular interest in the present context are processes where two   photons, 
emitted from the scattered electron and positron, fuse to
create quark-antiquark pairs, see Fig. 1c for the diagram describing the amplitude. 
Obviously, this amplitude is proportional to

   \hspace{3cm}   $T  \propto   (Q_{fred}^2 + Q_{fgreen}^2 + Q_{fblue}^2) / \sqrt{3}$\\ 
The corresponding  ratio $R_2$ of the cross section for photon-photon fusion 
into a specific  quark-antiquark pair  $f \bar{f} $ over that for
the fusion into $\mu ^+ \mu ^-$, again neglecting QCD radiative corrections and quark mass dependences 
far above threshold is

    \hspace{5cm}   $R_2 = 3 \langle Q_f ^2 \rangle ^2$ \\
With the SM assumption of fractional electric quark charges 
$R_2$ equals $ 3 (4/9)^2 = 16/27$ for up quarks
and $1/27$ for down quarks. 
The HN  scheme yields $3(2/3)^2 = 4/3$ for the fusion into up quarks and $R_2=1/3$ for down quarks. 
The differences (a factor 9/4 for up quarks and 9 for down quark, for the ratio of the HN
over SM cross section) are striking, in particular for the down-like quarks.

At LEP, the production of charm-anticharm and bottom-antibottom  
by photon-photon fusion has been studied, see \cite{L32005}. 
It has been observed that indeed the measured cross section for $b \bar{b}$ production 
is significantly  larger  than the expected (SM) cross section. 
Ferreira \cite{Ferreira} has proposed to adopt the theoretical calculations for the production by
the direct (unresolved) photon-photon fusion to charm and bottom quarks %, as shown in Fig. 1c, 
and to multiply them with the factors 9/4 and 9, respectively. 
Good agreement between prediction and experiment is achieved for this calculation. 

This is considered as a further  support for the Han-Nambu scheme.

\section{Conclusion} 
%The Standard Model has so far adopted the assumption  that the observed fractional electric charge 
%of each quark flavor (averaged over the color) coincides with the color-individual electric charge. 
%However, 
The assumption that quarks have integer electric charge according to the Han-Nambu scheme 
largely simplifies the   orthogonality condition between photon and $Z$ boson,  in  leading order.  
It allows a straight-forward  derivation of  the Weinberg angle, 
based on all  fermions of the three known families. 
Surprisingly, the result  is almost independent  of the mass sector of the Standard Model. 
The obtained value of $sin^2(\theta _W) = 1/4$ 
deviates  only by 8$\%$ from the  quoted present best theory-corrected experimental value.
%This near coincidence cannot be accidental.   
 
\noindent \textbf{Acknowledgment}\\
\noindent The support of my colleagues 
at the Cluster Origin and Structure of the Universe, Munich, 
%and at CERN, Geneva, 
is gratefully acknowledged. In particular, I thank    
Dorothee Schaile, Stephan Paul and Otmar Biebel.  

%P.M.Ferreira,  “Can we build a sensible theory with broken charge and colour symmetries”, hep-ph/0210024  (March 2002)
% [Kniel2000]                   B.A.Kniel et al., NPB582(2000)514  Fragmentation functions
%[LeeKim1978]               somewhat strange paper on Deep inelastic COMPTON scattering
%[Nisius2000]            R.Nisius, Phys.Rep. 332 (2000) 165-317 

\end{document}